\documentclass{emulateapj}

\usepackage{lscape}

\newcommand{\minusone}{$^{-1}$}
\newcommand{\kms}{km~s$^{-1}$}
\newcommand{\kmsm}{km~s$^{-1}$~Mpc$^{-1}$}
\newcommand{\hicont}{$\langle {\rm HI} \rangle$}
\newcommand{\Ha}{$\rm H\alpha$}
\newcommand{\rmax}{$R_{\rm max}$}
\newcommand{\ropt}{$R_{\rm opt}$}

\newcommand{\rd}{$r_{\rm d}$}

\newcommand{\rpe}{$r_{\rm PE}$}
\newcommand{\vpe}{$V_{\rm PE}$}
\newcommand{\wpe}{$W_{\rm PE}$}

\newcommand{\about}{$\sim$}

\slugcomment{Accepted for publication by the Astronomical Journal}

\shorttitle{Rotational Widths for Use in the TF Relation.}
\shortauthors{Catinella et al.}

\begin{document}

\title{Rotational Widths for Use in the Tully-Fisher Relation. I. Long-slit
  Spectroscopic Data}

\author{Barbara Catinella\altaffilmark{1}, Martha P. Haynes\altaffilmark{2}, 
\& Riccardo Giovanelli\altaffilmark{2}}

\altaffiltext{1}{National Astronomy and Ionosphere Center, Arecibo Observatory, 
HC3 Box 53995, Arecibo, PR 00612, USA; bcatinel@naic.edu.
The National Astronomy and Ionosphere Center is operated by Cornell University 
under a cooperative agreement with the National Science Foundation.}
\altaffiltext{2}{Center for Radiophysics and Space Research and National Astronomy and 
Ionosphere Center, Cornell University, Ithaca, NY 14853, USA; haynes@astro.cornell.edu, riccardo@astro.cornell.edu.}

\begin{abstract}
We present new long-slit \Ha\ spectroscopy for 403 non-interacting
spiral galaxies, obtained at the Palomar Observatory 5 m Hale telescope,
which is used to derive well-sampled optical rotation curves. Because
many of the galaxies show optical emission features which are significantly
extended along the spectrograph slit, a technique was devised 
to separate and subtract the night sky lines from the galaxy emission.
We exploit a functional fit to the rotation curve to identify its center of
symmetry; this method minimizes the asymmetry in the final, folded
rotation curve. We derive rotational
widths using both velocity histograms and the {\it Polyex} model fit. The final
rotational width is measured at a radius containing 83\% of the total light as derived from
I-band images. In addition to presenting the new data, we use a large sample of 742
galaxies for which both optical long-slit and radio HI line spectroscopy
are available to investigate the relation between the HI content of the
disks and the extent of their rotation curves. Our results show that the
correlation between those quantities, which is well-established
in the case of HI-poor galaxies in clusters, is present also in
HI-normal objects: for a given optical size, star formation can be
traced further out in the disks of galaxies with larger HI mass.
\end{abstract}


\section{Introduction}

Accurately determined rotational velocities of spiral galaxies are
important tools of extragalactic astronomy and cosmology. Among other
applications, they serve to derive masses and redshift-independent
distances of galaxies, to characterize the peculiar velocity field in
the local universe, to probe the impact of tidal and collisional forces
and to derive mass-to-light ratios and their evolution. Of particular note,
the accuracy of the rotational velocity measurement is the single dominant
contributor to the error budget of the Tully-Fisher \citep[TF;][]{tf77}
relation, beyond the intrinsic scatter \citep{gio97b}.
The full characterization of a galaxy's velocity field ideally requires
a two-dimensional map, such as can be obtained by means of HI synthesis,
Fabry-Perot or fiber bundles. Such maps can provide direct evidence
for departures in the velocity field from circular rotation and planarity
and can best allow the extraction of the contribution of circular rotation.
However, those techniques have become available relatively recently,
and they are still observationally costly (Fabry-Perot systems
do not allow for simultaneous sampling of the velocity channels) or
characterized by limited spatial dynamic range (fiber bundles).
For these reasons, the vast majority
of velocity width data have relied on simpler, albeit cruder, observations, 
e.g. one-dimensional
rotation curves (RC) $V(r)$ obtained with single-slit spectrographs placed
along a spiral disk's major axis or, even more commonly, with the zero-th
order approach of 21cm global (i.e., spatially integrated) line
profiles obtained with single radio dishes.
In practical situations such as wide area peculiar velocity studies exploiting
the TF relation, it is often necessary to combine rotational velocity
measures obtained from multiple methods. In such cases, it is imperative to
devise width measurement algorithms that are both robust and minimize technique-based
bias or systematic effects. This paper is the first of a series in which we
use a large compilation of both global HI line profiles and optical long-slit
spectra to minimize such systematics for TF applications and to derive a
statistical characterization of how disks rotate.

Most optical studies of rotation velocities in the low redshift universe
concentrate on the red part of the optical spectrum, made attractive by 
the presence of several nebular lines, most prominently \Ha\ at
6563\AA\ and [NII]$\lambda\lambda$6548,6584\AA.
The radial variation of the observed rotational
velocities -- the rotation curve -- offers insight into the velocity field but
must be interpreted in light of assumptions about dark matter components, 
spatial and kinematic disturbances and asymmetries, the impact of extinction,
and likely biases induced by differences in the global star formation history.
Several recent works have contributed significant compilations
of velocity widths based on optical long-slit spectra, including
\citet{mfb92}, \citet{mf96}, \citet[hereafter C97]{cou97},
\citet{vog95}, \citet{vog04a}, \citet{dal98}, and Dale et al. (1997, 1998, 1999a,b,c, 2001).
Rotational widths presented by those authors have been derived using several
different algorithms, based either on velocity histograms (Mathewson,
Vogt, and their collaborators) or on functional fits to the RCs (C97;
Dale and collaborators). The current work presents new data and widths based
on a measurement technique belonging to the second category. Discussion
of the systematics using a large compilation including all of the above referenced
datasets but with rotational widths remeasured on the digital spectra according to the 
technique presented here will be presented separately 
\citep[hereafter Paper II]{chg05}.

The principal issue is: given a RC, what specific measurement of a characteristic
velocity width best reflects a physical quantity tied to the galaxy's rotational velocity.
The circular velocity field of a galaxy can be calculated from the expression 
describing the gravitational potential
of the system, under the assumption of centrifugal equilibrium. In the
case of an infinitely thin disk, characterized by an exponential
surface density distribution with scale length \rd, it can be shown
\citep{fre70} that the resulting RC reaches a peak at 2.12 \rd, and
approaches the asymptotic Keplerian law at larger galactocentric
distances. In this simple case, one could identify the velocity width
with the value of the maximum rotation of the disk (or twice its value).
The observational discovery that most RCs are not falling, but rather
flat or even rising at the edge of the optical disk, raised the issue
of accounting for matter that was not detectable except for its
gravitational effects. The circular velocity field should be thus
calculated from the sum in quadrature of two terms, related to the
exponential disk and the dark matter halo; the effect of a stellar
bulge can be included similarly. 

For simplicity, the ``dark'' halo is usually assumed
to be spherical. Two common choices for the expression of its radial density
distribution are the so-called isothermal and NFW \citep{nfw97}
profiles, the latter derived from N-body simulations of structure
formation in hierarchically clustering universes. The observed RCs, or
at least their large-scale, smooth behavior (neglecting spiral streaming,
bar perturbations, disk warps or asymmetries, and so on), can be
successfully modeled using these components, and giving them
appropriate weights in order to optimize the results. The presence of
a dark matter halo complicates however the definition of rotational
width, since it should no longer be associated with the peak of the RC,
but the appropriate radial distance at which a characteristic width
measurement should be made is not obvious. 
While several authors measure rotational velocities at 2.15 \rd\
(e.g., \citealt{cy95} and C97), others prefer to do it at
the optical radius \ropt, i.e., the radius encompassing 83$\%$ of the
total integrated light of the galaxy (e.g., \citealt{pss96,gio97a}. 
For comparison, \ropt =3.2 \rd\ for an exponential disk). 
The sampling of the velocity fields at larger galactocentric distances 
offers important advantages in terms of smaller extinction and
shallower (if not null) velocity gradient, since both dust fraction
and RC slope decrease with increasing radius. 
For these and other reasons, further discussed here and in
Paper II, we adopt \ropt\ as reference radius for width measurements.
Our approach to determining rotational velocities
from long-slit RCs is described in detail in this work.
Rotational velocities obtained from optical long-slit spectra should, however,
be compared to radio widths derived from HI integrated line
profiles, using large samples of galaxies with both measurements available.
This is a key element to address the issue of a reference scale for
RC width measurements. Since HI emission can be traced out to radial distances that are
typically twice as large as those sampled by HII regions, HI widths
should provide, in fact, more reliable estimates of the maximum rotation of the disks.
Such comparison will be presented in Paper II.

This paper is organized as follows: We describe our sample selection
criteria, observation strategy, and data reduction process in Section 2. 
The RC extraction and folding technique are discussed in Sections 3
and 4. We illustrate our velocity width measurement method
in Section 5, and present our results in Section 6. The relation
between the distributions of the \Ha\ and HI line emissions in the
disks, which is relevant to the determination of velocity widths,
is investigated for a large overlap sample with both radio and optical spectroscopy 
in Section 7, and our conclusions are summarized in Section 8.

\section{Observations and Data Reduction}

The optical long-slit spectroscopy presented in this work was obtained at the Palomar Observatory
5 m Hale\footnote{
The Hale telescope at the Palomar Observatory is operated by the California Institute 
of Technology (Caltech) under cooperative agreement with Cornell University 
and the Jet Propulsion Laboratory.}
telescope. A total of 20 nights (19 of which had weather conditions adequate for 
spectroscopy) were allocated over the course of five observing runs.
To minimize overheads and maximize throughput, spectra were obtained using only the red
camera of the Double Spectrograph \citep{oke82} at the Cassegrain focus (f/15.7). 
The combination of the 1200 lines mm\minusone\ grating and a 2\arcsec\ wide slit yielded a dispersion of 
0.65 \AA\ pixel\minusone; the spatial scale of the CCD was 0.468\arcsec\ pixel\minusone. 
The grating angle was set in order to obtain a spectral coverage of 6530-7180 \AA, 
allowing detection of the \Ha\ line up to $cz \simeq$ 28,000 \kms. In addition,
a small number of spectra were acquired with similar instrumental setup during other observing 
runs by members of our group. A few galaxies have been observed more than once
to check the consistency among different runs. Of the 532 galaxies targeted, 488 show
\Ha\ emission (12 of the remaining ones show \Ha\ in absorption; the
other 32 have continuum but no line emission within the observed
wavelength range). The final number of galaxies with useful, extended RCs is 403.

The sample targeted for the long-slit observations was extracted from
a larger set of spirals identified as good TF candidates for inclusion in a 
photometric I-band survey carried out by M.P.H., R.G., and
collaborators as a part of an effort to map the peculiar velocity field in the local
Universe. Several datasets containing I-band photometry, optical RCs
derived from long-slit spectra, and global HI profiles all in digital
format contribute to what will be referred to as the SFI++ sample;
details of this compilation will be discussed more fully elsewhere
(Catinella 2005; Masters et al., in preparation).
The spectroscopic sample presented here includes undisturbed, inclined spirals with
apparent diameter greater than 0.6\arcmin, as estimated from a quick-look examination of the 
I-band images. The targets were identified principally to
provide adequate TF target sampling in the sky area just south of the
celestial equator, $-15^{\circ}\leq
\delta \leq 0^{\circ}$, and to provide optical rotation widths for
objects for which HI global profiles were not available for one of various reasons.
The resulting sample for the \Ha\ spectroscopy was heterogeneous,
and included three types of targets:
(a) galaxies with no spectroscopy (optical or radio) available;
(b) galaxies with problematic HI profiles (due to confusion within the radiotelescope beam,
or with irregular, asymmetric profiles, where a velocity width could not be clearly
determined), and
(c) a minority of nearby galaxies with apparent sizes larger than the spectrograph slit,
where observations obtained at multiple offset positions along the
major axis were required to trace the full extent of the velocity field.
A few additional targets (14 galaxies) for which I-band photometry was
not available were observed to fill in gaps in our observing schedule.

Due to the heterogeneous composition of the observational sample, we adopted a slightly different
observing strategy depending on the category (a, b, or c in the previous paragraph)
to which each target belonged.
In the absence of previously available spectroscopic information (the majority of the cases),
we performed a first, 5 minute integration on the target. A quick
inspection of the spectrum allowed us to check for the presence of
extended galaxy line emission, 
and to estimate the strength of the spectral lines. If the observation was deemed 
useful for TF purposes, a second integration long enough to adequately sample the outer disk 
regions (typically 10-15 minutes) was performed. For some other targets, prior information
on the strength of HI line emission suggested that \Ha\ would be detectable; in such cases,
we performed a 10 minute integration.
Finally, in those cases where the 128\arcsec -long slit did not sample the full extent of the RC,
we acquired one or more ``offset'' spectra, displaced by multiples of 45\arcsec\ from the 
central position, along the major axis of the galaxy. This technique was particularly designed
to insure that the outer parts of large objects are adequately sampled.\\

The spectra have been reduced using standard and customized IRAF\footnote{
IRAF (Image Reduction and Analysis Facility) is distributed by the National Optical Astronomy 
Observatory, which is operated by the Association of Universities for Research in Astronomy 
(AURA) under cooperative agreement with the National Science Foundation.}
tasks. 
The standard reduction procedure includes the following steps: bias subtraction, flat-fielding, 
cosmic ray cleaning, line-curvature correction (modeled from the night sky emission lines; 
the ``S-distortion'' was found not to be significant and thus ignored), and wavelength 
calibration (using the sky lines). Multiple exposures of the same galaxy at the same slit position 
were co-added, sky lines and galactic continuum were fit with polynomials and then subtracted
from the spectra. In the remainder of this section, we will discuss the sky subtraction 
in detail, focusing on the case of galaxies with line emission significantly extended relative 
to the size of the slit.

\subsection{A New Technique for Sky Line Subtraction}

The {\em standard} sky subtraction technique requires the
identification of {\em signal-free} regions on 
both sides of the galaxy line emission, along the spatial direction;
for each spectral channel, the sky lines in those narrow strips are
then interpolated with a low-order (2 or 3, typically) polynomial, 
and subtracted from the spectrum.
Such a procedure is clearly not applicable if the galaxy line 
emission fills the spectrograph slit. In particular, this problem affects 73 galaxies in our 
sample which are sufficiently extended to require offset observations. We thus developed 
a modification of the standard technique to deal with this issue.

\begin{figure}
\plotone{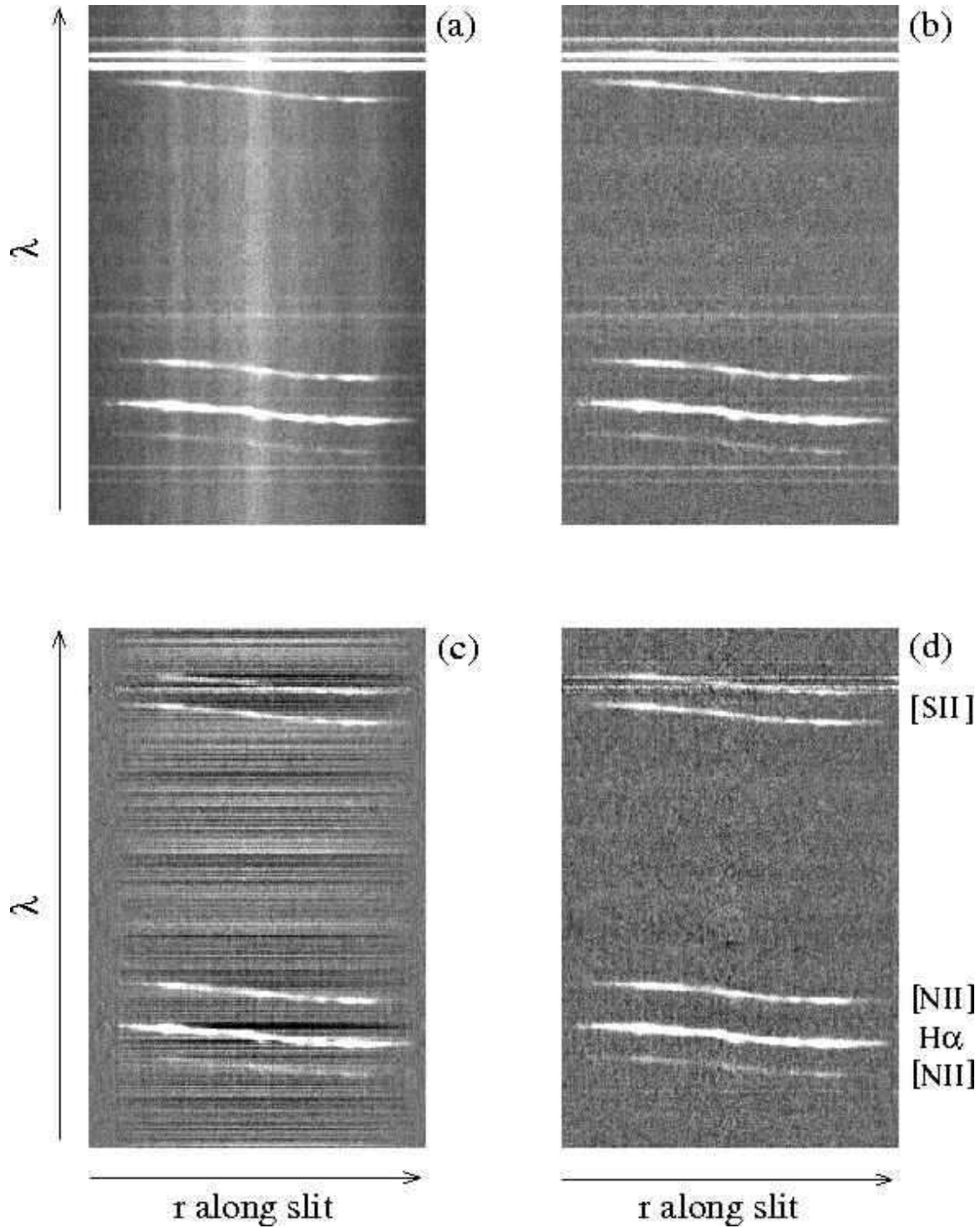}
\caption[Sky Line Subtraction from Optical Spectra]
{Demonstration of the template sky line subtraction process for MCG-2-31-017, a galaxy with line 
emission which fills the slit (we show here only the region with galaxy emission lines; the 
wavelength range runs approximately from 6610 to 6860 \AA).
Top panels: spectrum before (a) and after (b) galaxy continuum subtraction.
Bottom panels: comparison between standard (c) and {\em template} (d) sky subtraction.
In (d) the sky lines have been removed by subtracting a {\em sky line template} (see text).
Different galaxies require, in general, different templates.
\label{sky_sub}}
\end{figure}

\begin{figure}
\plotone{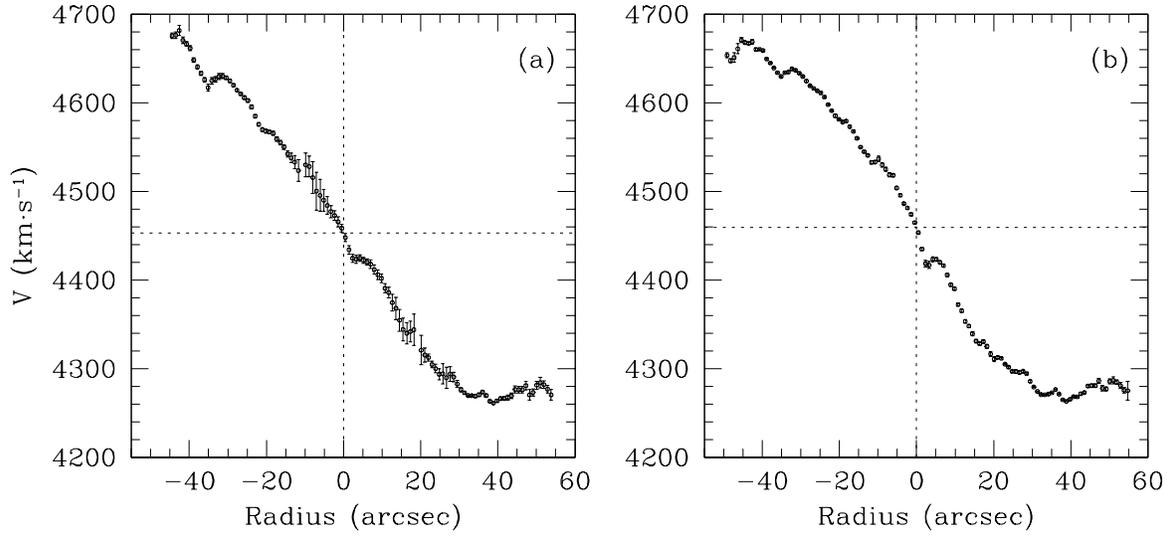}
\caption[Standard and Template Sky Subtraction]
{(a) \Ha\ RC of MCG-2-31-017, extracted from the sky line-subtracted spectrum 
in Figure~\ref{sky_sub}c. The dotted lines mark the position of the kinematic center (see \S~\ref{s_folding}, 
and Figure~\ref{folding}).(b) Same as (a), for
Figure~\ref{sky_sub}d. The large residuals present in
Figure~\ref{sky_sub}c result in larger error bars in (a) compared to (b).
\label{templ_rc}}
\end{figure}

We illustrate our technique (which will be referred to as {\em template} sky subtraction)
using the galaxy MCG-2-31-017 (AGC 520006) as an example. MCG-2-31-017 is an Scd spiral, characterized by 
weak continuum and strong line emission; its heliocentric velocity is 4460 \kms.
Figure~\ref{sky_sub}a shows the portion of its spectrum that includes the galaxy emission lines 
(before continuum and sky subtraction). The horizontal and vertical axes correspond to
spatial position along the slit and dispersion, respectively. The \Ha\ line is the most prominent 
feature in the lower half of the spectrum. The lack of signal-free regions at the edges of the 
slit leads to a very poor sky line subtraction, when the standard procedure is applied 
(Figure~\ref{sky_sub}c; the galaxy continuum emission has also been removed). 
The use of a few pixels on both sides of each row of Figure~\ref{sky_sub}a 
cannot, in fact, constrain the shape of the sky and continuum
emission, except at the very edges of the slit, and thus the sky
subtraction leaves large residuals across the spectrum (horizontal
chop in Figure~\ref{sky_sub}c). The result for the sky line-free
region would significantly improve if the continuum emission was removed
first (i.e., if the sky subtraction was applied to
Figure~\ref{sky_sub}b instead of \ref{sky_sub}a), but the residuals in 
the region of interest for RC extraction would be unchanged.
This result can be greatly improved by subtracting a sky line {\em template} 
spectrum, specifically constructed for MCG-2-31-017, from the continuum-free spectrum shown in
Figure~\ref{sky_sub}b. The sky template is a spectrum that contains
only sky lines, with positions and intensities that match those in
Figure~\ref{sky_sub}b as closely as possible, at least near the \Ha\
line. It is obtained by taking the median of the continuum-subtracted spectra of three or 
more {\em reference} galaxies, showing no line emission in the redshift region where the \Ha\ line
of the target galaxy lies. Each reference spectrum must be precisely
aligned with the target spectrum to match the positions of its sky
lines (with an accuracy of a few tenths of a pixel). Finally, the
intensities of the template sky lines are optimized for the target
spectrum by multiplying the template by a small numeric factor.
Such factor is chosen upon examination of the residuals of the
template subtraction near the \Ha\ line. If the residuals are positive
for some sky lines and negative for others, the scaling factor used is that 
providing the most reliable RC (in this sense, a sky line residual
running across the brightest part of the \Ha\ line does not affect the
RC, whereas one that overlaps with the low signal-to-noise parts does).
The identification of suitable reference galaxies can be very challenging, because not 
only the intensity of each sky line changes (depending on integration time and atmospheric conditions), 
but also the relative intensity of different sky lines can vary significantly over the course of 
the same night. More details on the choice of reference spectra can be found in \citet{cat05}.
The result of the template subtraction for MCG-2-31-017 is shown in Figure~\ref{sky_sub}d:
our method has substantially improved the quality of the sky
subtraction. 
Figures~\ref{templ_rc}a and \ref{templ_rc}b show the RCs extracted from the spectra in 
Figures~\ref{sky_sub}c and \ref{sky_sub}d, respectively. The dotted lines indicate the center of 
symmetry of the RC, identified by the average of the 90th and 10th percentile velocities (kinematic
center, discussed in \S~\ref{s_folding}). The error bars are obtained from Gaussian fits to
the \Ha\ line profile at each spatial position, as explained in \S~\ref{s_rcextr}.
A close inspection of Figure~\ref{templ_rc} shows that the RC in (b)
is less noisy and traces slightly farther out than that in (a). 
The number of data points (112 and 104, respectively) affects the values
of the percentile velocities used to identify the kinematic center,
and accounts for the small offset (6 \kms) between the central velocities in the two cases. 

In general, different templates are required for different objects, nights, or
even different emission lines of the same galaxy, making this sky subtraction technique
very time-consuming (unless the sky lines are fairly stable over a
given night, in which case the same template might be used for multiple target spectra).
Whether the result is worth the effort involved depends on the purpose of the observation and on the 
characteristics of the emission of the galaxy considered. In fact, the
standard and template sky subtraction provide essentially the same RC
width for MCG-2-31-017, but for galaxies with fainter \Ha\ emission,
or for purpose of studying the details of the RC in its outer parts especially, this 
template fitting technique can produce a more robust measurement than the standard approach.\\

It is interesting to mention another approach to a similar
sky-subtraction issue, developed by \citet[][hereafter B05]{ber05} for
integral-field fiber spectroscopy. 
The subtraction of night sky emission for multifiber data has been a
long-standing problem, usually dealt with by adopting observing
techniques in which the telescope is nodded between targets and
adjacent sky positions. This strategy is successful but not very efficient,
since only half of the observing time is spent on the targets. An
alternative solution based on a different sky subtraction algorithm
has been recently proposed and tested on spectra obtained with
SparsePak, a formatted fiber array optimized for spectroscopy of low
surface brightness galaxies, described in \citet{ber04}. 
Following the approach explained in detail in B05, the
wavelength-calibrated and rectified multifiber spectra obtained from
each observation are arranged into a two-dimensional image format,
sorted by position along the slit (with the interfiber gaps taken out). 
The result is analogous to, and processed as, a long-slit spectrum.
As Figure 14 of B05 shows, the problem is similar to that discussed in
this section: the source line emission spans the whole extent of the
slit, making the subtraction of the sky lines very difficult.
In the case of the SparsePak observations, 7 of the 82 fibers were
assigned to sky. Sky subtraction schemes that rely on a very small
number of sky {\em pixels}, however, yield noisy results, similar to
those shown in Figure~\ref{templ_rc}c for our dataset. A significant
improvement is obtained by using all the spatial channels (fibers): 
the spectrum is continuum-subtracted first (as we do for our spectra
before template subtraction), and a low-order polynomial is then fit
to each spectral channel, with a clipping algorithm that rejects
spatial channels with source flux.
This algorithm can be applied to continuum-subtracted long-slit
spectra, provided that (a) there are enough source-line-free pixels at
each wavelength to constrain the fit to the sky line profile, and (b)
the shape of the sky line profiles is symmetric with respect to
the center of the slit (so that the sky line fit over half slit fairly
matches the other half). The second condition is certainly met by our
Palomar spectra. In fact, for galaxies with line emission spilling
over {\em one side only} of the slit, we removed the sky lines by
fitting a {\em first-order} polynomial using only the signal-free side
of the spectrum (if there were enough pixels to do so). When this
method is applicable, it produces final spectra that are generally cleaner 
than those obtained via template subtraction.

In conclusion, both B05 and our template approaches can provide a
significantly improved sky subtraction for long-slit spectra where the
galaxy emission fills the slit. Which method should be used depends on
the characteristics of the source emission -- whereas B05's algorithm
avoids all the complications associated with the construction of a sky
template (the choice of suitable reference galaxies, image registration, and
intensity rescaling), its application requires a large enough number
of signal-free pixels at each wavelength to constrain the sky line fits.

\section{Rotation Curve Extraction}\label{s_rcextr}

Once continuum and sky line subtracted, the spectra are ready for RC extraction. 
Since the typical seeing at the Palomar Observatory is $\sim$1\arcsec-2\arcsec, and 
the CCD spatial scale used was \about 0.5\arcsec~pixel\minusone, the RCs are clearly
oversampled. Therefore, we Hanning smoothed the spectra and recorded the peak
value of the velocity (from a Gaussian fit across the \Ha\ line) for every other pixel.
The errors on the peak velocities were obtained through a Monte-Carlo resampling 
technique (an algorithm used by non-linear fitting {\em tasks} in IRAF
to estimate coefficient errors): each section across the spatial axis of the Hanning-smoothed spectrum was
replicated 15 times, each point being replaced by its value from the Gaussian fit
plus noise with dispersion given by the RMS of the fit. A Gaussian fit was performed
for each replication, and the dispersion of the fitted peak values used to estimate the 
velocity errors. All the points of the extracted RC were individually
inspected, with particular attention to those belonging to low
signal-to-noise regions, and only those with reliable Gaussian fits were kept.
The profile of the \Ha\ line at each spatial position is usually well
approximated by a Gaussian, with the possible exception of the
innermost parts of the galaxies, where non-circular motions sometimes cause
secondary peaks and deviations from the Gaussian shape 
(however such regions are typically those with the strongest \Ha\
emission, and therefore the positions of the profile peaks are well determined).

RCs with strong continuum emission or central bars, which mostly arise in earlier type (Sa-Sb)
spirals, often show \Ha\ absorption in the nuclear regions, due to the strong Balmer 
absorption features of A-type stars.
In these cases, we used another emission line, typically [NII]$\lambda 6584$ \AA\, if available, to 
fill in the central gaps of the \Ha\ RC. 
The [NII]$\lambda 6584$ \AA\ and \Ha\ RCs were extracted separately, matched, and the 
[NII]$\lambda 6584$ \AA\ data points kept only if lying within \Ha\ gaps. 
We applied a [NII]$\lambda 6584$ \AA\ patch to approximately 15$\%$ of our \Ha\ RCs.

When offset observations of the same galaxy (at the same slit position angle) were
available, the RCs separately extracted from each spectrum were combined by matching
their common features. In the region where the RCs overlap, we did not average the data points, 
but rather keep the ones with higher signal-to-noise ratio. As an aside, we point out that 
since offset observations farther from the galaxy center were typically integrated for a 
longer time, a stronger feature (i.e., smaller error bars) in the combined RC does not 
necessarily represent a stronger intensity of the underlying HII regions.

\section{Rotation Curve Folding}\label{s_folding}

\begin{figure}
\epsscale{0.85}
\plotone{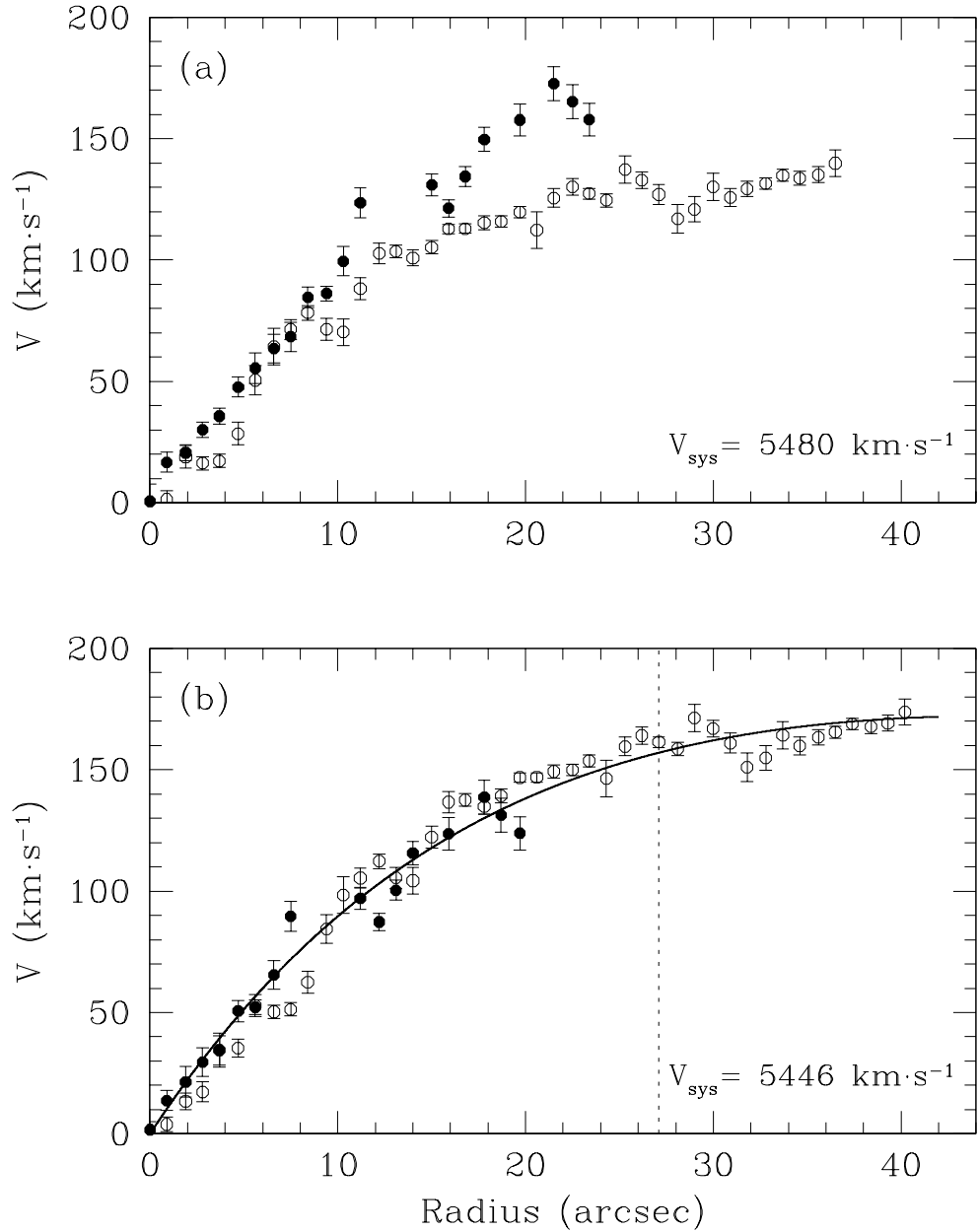}
\caption[Comparison of RC Folding Techniques]
{Comparison between kinematic (a) and URC (b) folding 
techniques for MCG-3-05-008. In (a) the RC center is identified by the data point with velocity 
closest to $V_{\rm sys}=(V_{10}+V_{90})/2$, where $V_{\rm N}$ is the Nth percentile velocity.
In (b) the center of symmetry has been determined by fitting a URC model to the unfolded RC. 
The best-fitting URC (not shown) indicates an offset of $-$34 \kms\ with respect to the kinematic center 
(no offset along the spatial direction), and provides a significantly better folding.
The solid line in (b) shows the Polyex fit (Equation~\ref{widths:eq_polyex}); the dotted line is 
at the optical radius, where the velocity width is measured. 
The RC is {\em not} deprojected to an edge-on view.
\label{folding}}
\end{figure}

In order both to measure the velocity width using our algorithm (see
\S~\ref{s_widths}) and to determine the systemic recessional velocity,
the rotation curves must be folded about a central origin, typically
chosen as either the {\em photometric} center-of-light, or the {\em kinematic} center. 
The definition of the latter itself varies; one common method identifies
the folding center with the velocity closest to the average 
of $V_{10}$ and $V_{90}$, the 10th and
90th percentile points of the RC velocity histogram (notice that since there are
no spatial interpolations involved, the two sides of the folded RC are
sampled on the same spatial grid). 
As shown by \citet{dal97}, kinematic centering appears preferable in most cases,
because it avoids misidentification of the photometric center due to bars or extinction
within the galaxy.
Moreover, the estimate of the light peak position is very poor in galaxies with a weak or 
nearly absent continuum emission (16$\%$ of our sample), and in galaxies with nuclear regions 
dominated by \Ha\ absorption (15$\%$ of our sample).
On the other hand, if a RC is significantly asymmetric or unevenly sampled (as in the case 
of galaxies that required offset observations), the kinematic centering will also
provide an unsatisfactory folding. This is shown in Figure~\ref{folding}a for MCG-3-05-008
(AGC 410269), a galaxy with a rising RC, where the \Ha\ emission is traced farther out 
in one of the arms.

A better result can be obtained making use of the detailed shape of
the RC by fitting an anti-symmetric function
to the unfolded RC and leaving the spatial position and velocity of the
center as free parameters. For its convenience and in order to minimize the number of
fitting parameters, we adopt the {\em Universal Rotation Curve} (URC)
description proposed by \citet{pss96} which incorporates contributions both from
a stellar exponential thin disk and a spherical dark matter halo.
Although inadequate to model individual RCs as already pointed out
elsewhere \citep[e.g., C97;][]{wil99}, the URC provides a useful
parameterization of the {\em average} RC of a galaxy in terms of its
luminosity and optical radius \ropt. In particular, the URC description approximates 
the overall shape of a RC adequately for the purpose of determining its center of symmetry.

In order to apply the URC folding method, we make use of the detailed SFI++ I-band photometric
data available to us, assigning to each galaxy its corresponding URC model based
on its I-band luminosity, \ropt, and inclination, and leave
a scale factor and a small offset from the kinematic center, in both
radial and velocity directions, as free parameters determined
via a non-linear least square fit to the RC data points.
Figure~\ref{folding}b demonstrates that this technique ({\em URC
folding}) is remarkably more successful than the simple kinematic one based
on the histogram method, shown
in (a), at folding asymmetric RCs. As expected, for symmetric RCs the two
folding methods give essentially the same result.
As discussed in \S~\ref{s_widths}, we measure the rotational velocity
of a disk galaxy at \ropt\ from a functional fit ({\em Polyex} model,
also shown in Figure~\ref{folding}b) to its folded RC.
It should be noted that the RC folding is not strictly necessary for
this type of width measurement, but the identification of the RC
center cannot be avoided. In fact, other authors prefer to add two
parameters to the expression used to model a RC, solving for
its shape and center at the same time (e.g., C97).

In order to assess the comparative validity of the kinematic and URC folding techniques on a
quantitative basis, we make use of the larger SFI++ sample and test their respective
ability to minimize the asymmetry of the folded RCs. 
As a measure of asymmetry, we adopt the definition given by 
\citet{dal03}:
\begin{eqnarray}
    {\rm Asymmetry} = \sum
        \frac{||V(R)|-|V(-R)||}{\sqrt{\sigma^2(R)+\sigma^2(-R)}} \nonumber \\
	\times \left[\frac{1}{2} \sum \frac{|V(R)|+|V(-R)|}{\sqrt{\sigma^2(R)+\sigma^2(-R)}}\right]^{-1}
\label{eq_asymm}
\end{eqnarray}
\noindent
where $V(R)$ and $\sigma(R)$ are velocity and corresponding uncertainty at the radial 
position $R$, and the summation symbols refer to radial distances for which {\em both arms} are 
sampled (unpaired RC data points are excluded). The quantity defined
in Equation~\ref{eq_asymm} simply measures the area
between the two folded RC arms, normalized by the average area that they subtend.
The calculation of the asymmetry index is slightly more complicated 
for the URC-folded RCs, because in general the two arms are {\em not}
sampled on the same spatial grid (due to a small spatial offset from
the kinematic center). We therefore calculate two indices for each RC by pairing
the points on one arm with those obtained by interpolation of the
other arm at the same spatial positions, and take their average as our
measure of asymmetry. 

\begin{figure}
\plotone{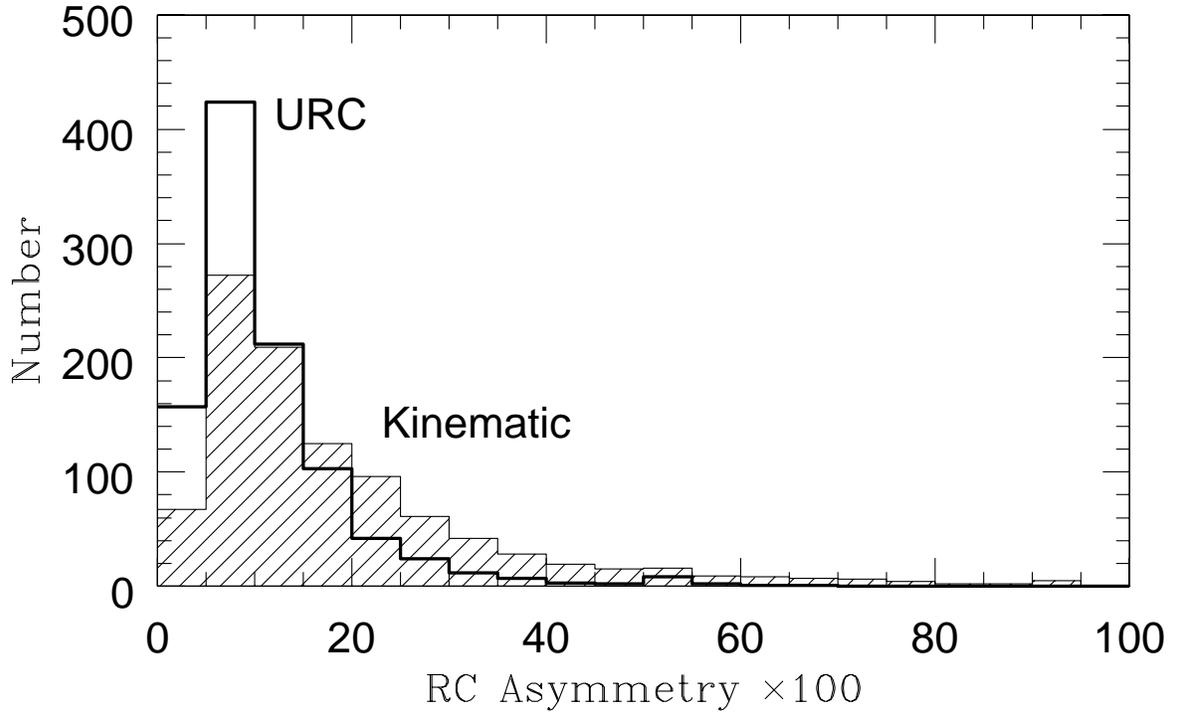}
\caption[RC Asymmetry Index Distributions]
{Distributions of the asymmetry indices for RCs folded according to the URC (solid histogram) 
or the kinematic (hatched histogram) techniques (see Figure~\ref{folding}). The two indices have been
calculated for the optical RCs in our digital archive, for which velocity errors were
also available (999 objects). The URC distribution is more centrally concentrated, and does not have a 
significant tail of high ``RC asymmetry'', showing that this algorithm is more effective at minimizing 
the overall asymmetry of the folded RCs.
\label{rcasymm}}
\end{figure}

The distributions of the URC and kinematic asymmetry indices of the
999 optical RCs with available velocity error information in our
digital archive are presented in Figure~\ref{rcasymm}. The histogram
obtained for the URC folding (solid) is more centrally concentrated
than the one corresponding to kinematic folding (hatched), indicating
that the first technique is more effective at minimizing the overall
asymmetry of the RCs. The mean asymmetry indices for the URC and
kinematic cases are 11.3$\pm$0.3\% and 19.4$\pm$0.5\%,
respectively; for comparison, \citet[hereafter D01]{dal01} obtained an average
asymmetry of 12.5$\pm$1.0\% applying Equation~\ref{eq_asymm} to a
sample of field galaxies (in the foreground or background of
clusters) with RCs folded according to the kinematic histogram technique. 
Both D01 and our samples were selected for TF studies, have similar
morphological composition (dominated by normal late-type spirals, mostly Sc), 
and were obtained with similar instrument setup (most of the D01
spectroscopy was carried out with the Palomar 5m telescope);
however the characteristic redshift of our galaxies is smaller
(typically $z\lesssim 0.03$, as opposed to $0.02\lesssim z \lesssim
0.08$ for D01). The loss of spatial detail of the RCs of the more distant
galaxies probably accounts for the smaller asymmetry index obtained by D01
using the histogram technique. 
The folding of spatially well-sampled RCs should be done by
taking into account the RC shape. Folding algorithms that rely on velocity
histograms alone might in fact introduce spurious asymmetry in the folded RC due to non-optimal
matching of the two arms, as shown in Figure~\ref{folding} (the URC
and kinematic asymmetry indices for that RC are 0.17 and 0.74, respectively). 

\begin{figure}
\plotone{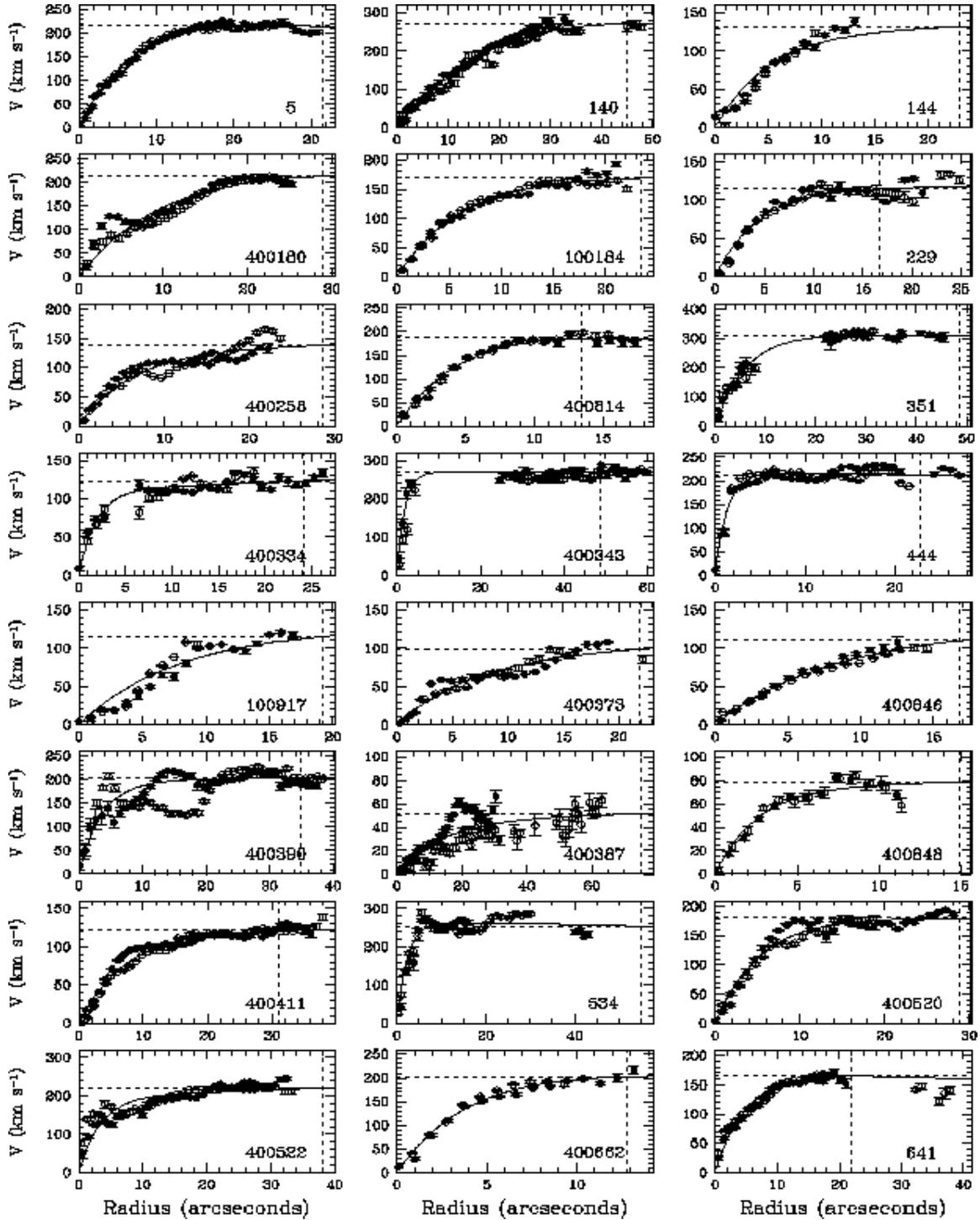}
\caption[Palomar \Ha\ Rotation Curves]
{\Ha\ rotation curves, Hanning-smoothed and recorded every other pixel to account 
for oversampling. The AGC identification code is given in each panel. 
In one case (AGC 12591) the RC has been extracted from the [NII]$\lambda 6584$ \AA\ emission line.
The error bars are formal errors from the Gaussian fits to the peak
velocities. The RCs are {\em not} deprojected to an edge-on view. 
The Polyex fit is shown by a solid line; the dotted lines indicate the
position of \ropt\ and the adopted half velocity width. We show here
only the first page.
\label{rc_1}}
\end{figure}

RC asymmetry indices have been derived in this section with the
limited intent of comparing the two folding techniques. 
It should be noted however that the asymmetry parameter itself could
be used to determine the folding, its minimum value identifying the
center of symmetry of the RC. For RCs that are well-sampled and traced
to similar extent on both sides, we expect this and our method to give
similar results, since fitting the URC is effectively a way of
minimizing the overall asymmetry of the RC. For other cases, however,
the URC technique has an important advantage: {\em all} the RC points
are used to determine the folding center and not only those lying at
radial distances that are sampled on both sides of the RC. 
In particular, our technique provides a satisfactory result for
galaxies (a) with patchy \Ha\ emission or (b) with the two sides
unevenly sampled (as it happens for our offset observations). 
While in case (b) the asymmetry index centering might
still yield a good result (even if using a smaller number of RC
points), there are examples of case (a) where the asymmetry index could not be
calculated (e.g., AGC 4271 and AGC 221928 in Figure~\ref{rc_1}; see
\S~\ref{s_data}).

A complete treatment of asymmetry in RCs, as well as a more general
comparison among folding techniques, are beyond the scope of this
paper, and will be addressed in a future work.

\section{RC Width Measurement}\label{s_widths}

As discussed e.g., by C97 and \citet{ray97},
algorithms to measure rotational velocities from RCs
can be divided into two principal categories: {\em histogram} and 
{\em model} widths. The former are obtained by collapsing the RCs into velocity 
histograms, and calculating the difference between $V_{\rm N}$ and $V_{\rm 100-N}$, where
$V_{\rm N}$ is the N-th percentile velocity (typically, N=10). This or
similar definitions have been adopted by, e.g., \citet{mfb92},
\citet{mf96}, \citet{vog95}, and \citet{vog04a}.
Model widths are measured at a fixed galactocentric distance (derived from photometry) 
from model fits to the RCs. A number of different fitting functions have been adopted
such as those discussed by C97; since they are all empirical, trade-offs between
simplicity and detail result. Authors who have published large
collections of model widths include C97, \citet{dal98}, and Dale et al. (1997, 1998, 1999a,b,c, 2001).
Histogram widths are easier to measure and do not require accompanying
photometry, but do not make use of the spatial information contained
in the RCs, and therefore, not surprisingly, they are inferior to the
model widths. A direct comparison between the two types of widths shows
in fact that the histogram ones are affected by biases that correlate with the
shape and extent of the RCs (Paper II). For example, histogram widths
derived from RCs with an extent \rmax $\leq$\ropt\ are typically
smaller than model widths measured at \ropt, the
difference being larger for smaller \rmax/\ropt\ ratios and steeper RCs
(for rising RCs, the average width difference is 20 \kms\ or larger,
depending on RC extent). 
When heterogeneous width measurements are used in a TF survey,
systematic trends such as the one described must be taken into account in
order to avoid biases in the analysis. For instance, a width difference
of 20 \kms\ translates into approximately a 0.2 mag offset on the TF
plane, or 500 \kms\ in peculiar velocity, at $cz\sim 5000$ \kms. 
Clearly, galaxies of different measures will be mixed in a statistical sample, and the 
net effect will be reduced. However, if the data sets are segregated
in sky coverage, spurious bulk flows of significant size can arise 
as a result.

The method adopted here to determine velocity widths involves
fitting a function \citep[{\em Polyex} model;][]{gh02} to the folded RC, 
and measuring the velocity from the fit, at the optical radius \ropt\ of the galaxy. 
The RC is folded using the URC technique described in \S~\ref{s_folding}. 
As its name suggests, the functional form of the Polyex model combines an exponential,
which follows the inner RC rise, and a polynomial, which reflects the outer RC slope:
\begin{equation}
        V_{\rm PE}(r) = V_0(1-e^{-r/r_{\rm PE}})(1+\alpha r/r_{\rm PE})
\label{widths:eq_polyex}
\end{equation}
\noindent
where  $V_0$, \rpe, and $\alpha$ determine the amplitude, the exponential scale of
the inner region, and the slope of the outer part of the RC, respectively.
The Polyex width is thus simply defined as \wpe =2~\vpe~(\ropt). 
An example of the Polyex width measurement technique for a representative galaxy is presented 
in Figure~\ref{folding}b. 

Considerable discussion has appeared in the literature concerning the spatial position
at which the RC width should be measured in light both of the relevant physics and
practical concerns about
fitting disk scale lengths and possible extinction effects. Following, e.g. C97,
many authors adopt the location of the peak rotational velocity for a pure exponential disk,
$r_{\rm disk}=2.15~h$, where $h$ is the disk scale
length. As mentioned previously, the inclusion of an extensive dark matter halo
eliminates such a predicted RC peak. Furthermore, several studies \citep{gh02, baes03, vg04}
have demonstrated clear evidence that extinction can significantly affect
the inner slopes of observed rotation
curves with the impact being dependent on the galaxy's intrinsic luminosity. For these
reasons, we elect to measure further out in the disk, at \ropt. It should
be noted that we measure \ropt\ as the radius containing 83\% of the total
light, not at a fixed number of scale lengths \citep{hay99}.
Further discussion of this issue is addressed in Paper II, where our width measurement technique 
is compared to others proposed in the literature. 

In those cases for which the RC does not reach
\ropt, application of this method requires an extrapolation of
the fit beyond the last measured point. As discussed further in Paper II, 
such extrapolation produces reliable velocity widths only when  
$\rm R_{max}/R_{opt} \geq 0.5$, where $\rm R_{max}$ is the RC extent 
(i.e., the maximum radial distance up to
which the folded RC is sampled). Because our observations have typically
been made with the Hale 5m telescope, they are of high sensitivity. In practice,
the Polyex function has enough flexibility to fit the vast majority 
(\about 94\%) of the individual RCs in the SFI++ dataset (including
those that are declining at large radii). 
Finally, it is worth mentioning that Equation~\ref{widths:eq_polyex}
is not based on any physical models of the velocity field of disk galaxies, 
but only represents a convenient, empirical expression to fit a large variety of RC
shapes. 

Observed velocity widths derived from RCs must be corrected for cosmological broadening
(to obtain the rest-frame velocities) and deprojected to an edge-on view:
\begin{equation}
        W^{\rm corr} = \frac{W^{\rm obs}}{(1+z)\,\sin i}
\label{widths:eq_wcorr}
\end{equation}
\noindent
where $z$ and $i$ are redshift and inclination to the line of sight of the galaxy, respectively.
Total errors on velocity widths typical of RC datasets considered here are of the
order of 20 \kms\ \citep{gio97a}.

\section{Data Presentation}\label{s_data}

Table~\ref{table:rcdata} lists the main parameters of the 403 galaxies for which 
final, extended (\rmax $\geq$ 0.5 \ropt) RCs have been obtained, ordered by increasing Right Ascension.
We show here only the first page to illustrate form and content\footnote{
Table~\ref{table:rcdata} is available in its entirety in the
electronic edition of the Journal.}.
The parameters listed are:

\noindent
{\bf Col. 1:} identification code in our private database, referred to as
the Arecibo General Catalog (AGC). If a galaxy is also listed
in the UGC catalog \citep{nil73}, its UGC and AGC codes are identical.\\
{\bf Col. 2:} NGC or IC designation, or other name, typically from the CGCG \citep{zwi68}
or MCG \citep{vor68} catalogs. CGCG numbers are listed in the form:
field number-ordinal number within the field; MCG designations are abbreviated to eight 
characters. \\
{\bf Cols. 3 and 4:} Right Ascension and Declination for the epoch J2000.0. Coordinates
have been obtained from the Digitized Sky Survey catalog and are accurate to within 2\arcsec. \\
{\bf Col. 5:} T, morphological type code according to the scheme in the RC3 catalog \citep{dev91};
code 1 corresponds to Sa, code 3 to Sb, code 5 to Sc, etc. A ``B'' following 
the type code indicates the presence of an identifiable bar. \\
{\bf Col. 6:} \ropt, optical radius in arcsec, corresponding to the isophote that includes 83$\%$ 
of the I-band flux. \\
{\bf Col. 7:} \rmax, maximum radial distance to which the folded RC is sampled, in arcsec. \\
{\bf Cols. 8--10:} parameters of the Polyex fit (see Equation \ref{widths:eq_polyex}) to the folded RC. \\
{\bf Col. 11:} flag that indicates the quality of the Polyex fit. The flag is set to 1
if the fit is deemed good or acceptable, and to 0 if the fit is poor. \\
{\bf Col. 12:} heliocentric velocity from the RC velocity histogram, $V_{80}=(V_{10}+V_{90})/2$
(\S \ref{s_folding}). \\
{\bf Col. 13:} velocity width from the RC velocity histogram, $W_{80}=V_{90}-V_{10}$, 
corrected for cosmological broadening only (\S \ref{s_widths}). \\
{\bf Col. 14:} \vpe, heliocentric velocity from the Polyex fit
and its uncertainty (estimated from the fit to the RC data points). \\
{\bf Col. 15:} \wpe, velocity width from the Polyex fit, corrected for cosmological broadening 
only (\S \ref{s_widths}), and its estimated uncertainty. 


The empty entries in the Table (columns 6--11, 14, and 15) refer to 44 galaxies for which
good quality photometry was not available (14 of which without an I-band image) and therefore, 
for which a Polyex width could not be measured.

In Figure \ref{rc_1}, we show the corresponding \Ha\ RCs\footnote{
A complete set is available in the electronic edition of the Journal.}, 
Hanning-smoothed and recorded every other pixel to account for oversampling (see \S \ref{s_rcextr}). 
The AGC identification code is given in each panel. In one case (AGC 12591)
the RC was obtained from the [NII]$\lambda 6584$ \AA\ emission line
(which was stronger and slightly more extended than \Ha).
The RCs are folded according to the URC technique explained in \S
\ref{s_folding}, and {\em not} deprojected to an edge-on view.
The solid line shows the Polyex fit, and the vertical dashed line indicates the position of \ropt, 
where the Polyex width is measured. The adopted half velocity width is indicated by the horizontal line. 
In those cases where a Polyex fit was not available due to lack of photometry, the folding was
done about the position of the kinematic center (see \S \ref{s_folding}).

\section{Dependence of RC Extent on HI Content}\label{s_hidef}

One of the factors that directly determine the observed rotational velocity of a galaxy 
is the radial extent within the disk out to which the adopted kinematic tracer is detectable.
For example, the velocity width of a galaxy with a rising RC, as measured from an HI 
global profile, will underestimate the true value if the distribution of the neutral
hydrogen is truncated (a situation that is typical for objects located 
in high density regions such as centers of clusters). RC widths obtained by means of velocity
histograms in spatially resolved data -- a method that closely resembles that used to measure 
radio widths from HI integrated line-profiles, although not
intensity-weighted -- will suffer the same bias if the tracer is truncated. 
The impact of a possible \Ha\ truncation on RC model widths is expected to be negligible, since the
measurement is performed at a fixed galactocentric distance, provided that the RC is extended
enough to allow for a reliable fit of its shape; nonetheless, such expectation 
should be carefully investigated.

Several studies have already provided evidences for the existence of a correlation
between \Ha\ and HI extents, most notably for galaxies in the Virgo
cluster \citep[e.g.,][]{kk98,rwk99}. 
This is also in agreement with previous findings of truncated 
star formation in HI deficient spirals. For instance \citet{hg86} found that HI poor 
galaxies in the Virgo cluster are redder than those of similar morphology but normal HI content, 
and concluded that the star formation has been quenched in those
objects. \citet{gav02} reached the same conclusion based on spectral energy 
distribution modeling of spectrophotometric data of a rich sample of
spirals in Virgo. More recently, \citet{vog04b} studied a sample of 296
infalling spirals selected from 18 nearby clusters (and 33 isolated
field galaxies for comparison) and showed that
the distributions of HI flux and of HII regions within the disks are clearly correlated. 

Here, we make use of the entire SFI++ dataset, which is not restricted to cluster galaxies,
to investigate the dependence of the extent of the \Ha\ emission
on HI content. The SFI++ compilation contains 715 galaxies which have
good quality optical RCs and HI flux measurements \citep{shg05}.
We follow the common practice of defining the comparative HI content, \hicont, as the difference, 
in logarithmic units, between the observed HI mass, $M^{\rm obs}_{\rm HI}$, and the value 
expected for an isolated galaxy of the same morphological type, $T^{\rm obs}$, and optical linear 
diameter, $D^{\rm obs}_{\rm opt}$ \citep{hg84,sol01}:
\begin{equation}
	\langle {\rm HI} \rangle = \langle \, \log M_{\rm HI}\,(T^{\rm obs},\,D^{\rm obs}_{\rm opt})\, \rangle 
		- \ \log M^{\rm obs}_{\rm HI}
\label{eq_hidef}
\end{equation}
\noindent
where $M_{\rm HI}$ is in solar units and $D_{\rm opt}$ in kpc. 
This definition is commonly referred to as the {\em HI deficiency} and
is used to identify truly HI poor objects. Note that in these units,
objects with \hicont $>+0.3$ are HI deficient by a factor of 2
relative to similar isolated galaxies, and those with \hicont $>+0.5$
by a factor of 3. It should also be noted that values of \hicont\ for
individual objects are very uncertain \citep{sol01} and therefore
this quantity should be calculated for a data set and interpreted in a
statistical fashion.
We used the equations in \citet[][hereafter SGH96]{sgh96} to calculate the observed HI mass from the 
corrected HI flux \citep[obtained from][]{shg05} and 
the linear diameter from the apparent blue visual size 
(using the velocity in the CMB frame to estimate the radial distance). 
In particular, we used the maximum likelihood linear regressions
of $\log (h^2 M_{\rm HI})$ on $\log (h D_{\rm opt})$ (where $h$ is the Hubble constant 
in units of 100 \kmsm) presented in Table 2 of SGH96 to 
calculate the expectation value of the HI mass in Equation \ref{eq_hidef}. 
The regression relations are given only for the morphological types Sa+Sab,
Sb, Sbc, Sc, and All; we thus assigned: S0a same as Sa,
Scd-Sd same as Sc, and other types same as All.

From among the galaxies with good quality RCs in SFI++, we extracted two subsets: those
with good HI-line emission profiles (715 objects) and those that were
marginally or non-detected in HI (27 objects). 
We calculated the HI content \hicont\ as defined above for the first
subset. For the second group, we obtained lower limits using upper
limits to the estimated HI mass, which is commonly 
calculated (e.g., SGH96) by assuming a rectangular HI emission profile of amplitude 
1.5 times the observed RMS noise per channel and a typical rotational width
(since a RC was available for each galaxy, we used the Polyex width).

\begin{figure}
\epsscale{0.8}
\plotone{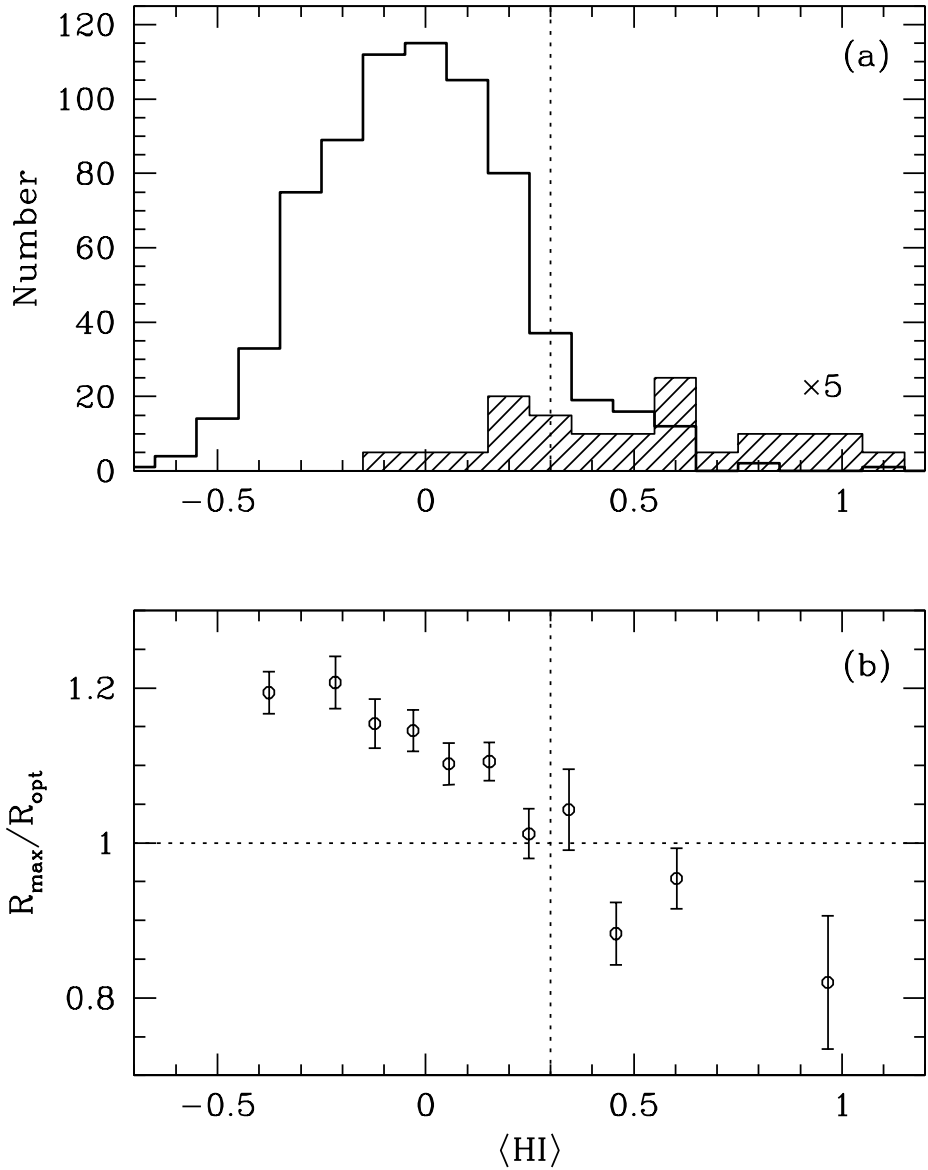}
\caption[Relation Between HI Deficiency and RC Extent]
{(a) HI content distribution for 715 SFI++ galaxies with optical RC and HI emission 
profile (solid histogram), and for 27 marginal or non-detections (hatched histogram; the numbers are 
multiplied by 5 for clarity). (b) Dependence of the RC extent on the HI content for all the objects in (a).
The points are running averages, each one including 100 ($x<0.2$), 59
($x=0.25$), 25 ($0.3<x<0.7$), or 8 ($x=0.97$) measurements; the error bars are Poissonian
errors on the mean.
\label{hidef}}
\end{figure}

The distribution of \hicont\ is shown in Figure~\ref{hidef}a for both detections (solid histogram)
and marginal or non-detections (hatched histogram; the numbers are multiplied by 5). 
The dotted line at \hicont=0.3 separates HI-poor (right) from HI-normal or HI-rich galaxies (left).
As noted previously \citep[e.g.,][]{vog04b} most of the truly HI deficient objects are not detected in HI, are 
found in cluster cores, and have truncated HI and \Ha\ disks. Such objects make up only a small fraction 
of the SFI++ RC sample. 
The significant new result is shown in Figure~\ref{hidef}b, where the RC extent 
(i.e., the maximum radial distance \rmax\ out to which the folded RC 
is sampled, relative to the optical radius \ropt) is plotted as a function of
\hicont\ for the overall sample. The correlation between these two
quantities, noted by previous authors in the case of HI-stripped
spirals in clusters (and confirmed by our sample, although the number of HI-poor objects is very small) 
{\em continues among the HI-normal galaxies}, in the sense that objects
with lower than average HI mass also have smaller than average \Ha\ disks.
Since both \hicont\ and \rmax/\ropt\ are normalized quantities, this
correlation is not a simple scaling relation.

These results are important for the interpretation of
systematic trends observed when velocity widths obtained from HI
integrated line profiles and long-slit optical RCs (using both
histogram and model approaches) are compared. As mentioned at the
beginning of this section, we would expect RC model widths to be
negligibly affected by \Ha\ truncation (as long as the extrapolation
to \ropt\ is deemed reliable). However the difference between HI and
Polyex widths does show a weak dependence on RC extent (normalized to
\ropt), in the sense that larger width differences are observed for
galaxies with \Ha\ emission traced farther out in the disk
\citep[Paper II;][]{cat05}. This bias is important for galaxies with 
steeply rising RCs, and can be explained by the correlation between \Ha\ and HI
disk sizes discussed here, which -- for rising RCs -- implies larger HI widths 
for galaxies with more extended \Ha\ emission. 
These results will be further discussed in Paper II, where the width
comparison will be presented in detail, and statistical corrections
will be determined with the purpose of making rotational velocities for TF work
technique-independent.

\section{Summary}

Here, we have presented new long-slit optical emission line RCs for 403 galaxies 
and have derived from them rotational widths of particular use 
for applications of the TF relation.  We described the RC extraction process and illustrated
a method to subtract the night sky lines from spectra which contain emission well extended
along the spectrograph slit. For galaxies with such extended emission,
this {\em template} sky subtraction technique 
yields RCs that are less noisy than those obtained with the standard
sky removal method, thus allowing one to sample the velocity field to a
larger spatial extent, especially when the galaxy line emission is faint.
Our approach to determine the center of symmetry about which
the RC folding is performed exploits use of the URC model and knowledge of
the galaxies photometric properties from I-band imaging observations. The measurement
of rotational velocity involves fitting an empirical function, the Polyex
model, to the folded RC and measuring the velocity from such fit, at \ropt,
the radius encompassing 83\% of the total I-band light. 

Because of its relevance to the measurement of rotational widths from
long-slit RCs, we have studied the relation
between HI content and RC extent for 742 galaxies in the SFI++
sample, for which both measurements are available (for 27 objects
marginally or non-detected in HI, lower limits for the HI content have
been derived). 
We found a clear correlation, in the sense that a larger than average HI content implies a
larger than average \Ha\ disk relative to the optical size. In
addition to confirming previous findings of \Ha\ truncation in HI-poor
cluster galaxies, our results show that the trend holds for HI-normal, field objects:
for a given optical size, star formation can be
traced further out in the disks of galaxies with larger HI mass.

The RC data presented complement other data for a large sample of several thousand 
disk galaxies for which I-band photometry and \Ha\ or HI-line spectroscopy 
are available. This combined SFI++ dataset is being used to investigate the
peculiar velocity field in the local universe via application of the TF method
as well as to perform detailed statistical studies on the nature of disk galaxies.
The detailed comparison between our and other RC width determinations
as well as the cross-calibration of optical and global HI line rotational
velocity measures will be
presented in Paper II. The same dataset is being used to derive 
average or {\em template} RCs in bins of galaxy luminosity,
which provide the basis for comparative
study of possible variations in the kinematic properties of disk
galaxies with increasing redshift \citep{cat05, cgh05}.

\acknowledgements

This research was supported by NSF grants AST-9900695, AST-0307396 and
by a NAIC pre-doctoral research grant at the Arecibo Observatory
to B.C. 
We wish to thank the referee Matthew Bershady for helpful comments on
the manuscript.
We acknowledge the use of NASA/IPAC Extragalactic 
Database (NED)\footnote{
NED is operated by the Jet Propulsion Laboratory, California
Institute of Technology, under contract with the National Aeronautics 
and Space Administration.
}
and NASA's SkyView facility\footnote{
SkyView was developed and is maintained under NASA ADP Grant
NAS5-32068 under the auspices of the High Energy Astrophysics Science
Archive Research Center (HEASARC) at the Goddard Space Flight Center
Laboratory of NASA.}.


\clearpage

\clearpage

\begin{landscape}
\begin{deluxetable}{rcrrlrrrrccrcrc}
\tabletypesize{\tiny}
\tablecaption{Palomar Rotation Curve Data. \label{table:rcdata}}
\tablewidth{0pt}
\tablehead{
\colhead{Galaxy} &  \colhead{Other} &
\colhead{$\alpha$} & \colhead{$\delta$} & 
\colhead{T} &
\colhead{\ropt} & 
\colhead{$R_{max}$} &
\colhead{$V_0$} & 
\colhead{\rpe} & 
\colhead{$\alpha$} & 
\colhead{Q} & 
\colhead{$V_{80}$} & 
\colhead{$W_{80}$} & 
\colhead{$V_{PE}$} & 
\colhead{$W_{PE}$}
\\
\colhead{} & \colhead{} & 
\colhead{J2000} & \colhead{J2000} &
\colhead{} &
\colhead{\arcsec} &
\colhead{\arcsec} &
\colhead{km s\minusone} &
\colhead{\arcsec} &
\colhead{} &
\colhead{} &
\colhead{km s\minusone} & \colhead{km s\minusone} &
\colhead{km s\minusone} & \colhead{km s\minusone}
\\
\colhead{(1)}&\colhead{(2)}&\colhead{(3)}&\colhead{(4)}&\colhead{(5)}&
\colhead{(6)}&\colhead{(7)}&\colhead{(8)}&\colhead{(9)}&\colhead{(10)}&
\colhead{(11)}&\colhead{(12)}&\colhead{(13)}&\colhead{(14)}&\colhead{(15)}
}
\startdata
     5 & 382-021  & 00 03 05.5 & $-$01 54 51  &   4B &     31.5  &     30.8  &  248.4 &  7.220 &   $-$0.0289 & 1   &  7303 &  420 &    7299$\pm$03  &  418$\pm$05 \\  
   140 & N  52    & 00 14 40.2 &   +18 34 55  &   3  &     45.1  &     47.8  &  374.0 & 20.576 &   $-$0.0823 & 1   &  5417 &  491 &    5414$\pm$08  &  535$\pm$11 \\  
   144 & 456-044  & 00 15 26.8 &   +16 14 05  &   4  &     22.9  &     13.1  &  124.9 &  5.000 & \phs 0.0150 & 0   &  5656 &  211 &    5640$\pm$18  &  259$\pm$25 \\  
400180 & N  64    & 00 17 30.2 & $-$06 49 29  &   4B &     28.9  &     25.2  &  249.9 & 10.695 &   $-$0.0321 & 1   &  7312 &  398 &    7312$\pm$09  &  416$\pm$12 \\  
100184 & I 1546   & 00 21 28.8 &   +22 30 24  &   4  &     23.4  &     22.0  &  172.5 &  6.031 & \phs 0.0000 & 1   &  5820 &  318 &    5820$\pm$04  &  331$\pm$06 \\  
   229 & 500-014  & 00 23 53.4 &   +28 19 57  &   3  &     16.6  &     24.8  &  117.9 &  4.452 & \phs 0.0003 & 1   &  7235 &  222 &    7234$\pm$04  &  225$\pm$06 \\  
400258 & M-202038 & 00 31 13.1 & $-$10 28 51  &   4B &     28.6  &     23.7  &  128.8 &  6.158 & \phs 0.0185 & 1   &  3603 &  256 &    3614$\pm$08  &  274$\pm$11 \\  
400814 & ...      & 00 31 56.7 & $-$10 15 08  &   4  &     13.5  &     17.4  &  218.5 &  4.730 &   $-$0.0331 & 1   & 11698 &  354 &   11701$\pm$05  &  359$\pm$07 \\  
   351 & I  34    & 00 35 36.5 &   +09 07 25  &   1B &     48.5  &     45.4  &  317.7 &  6.042 &   $-$0.0036 & 1   &  5339 &  618 &    5338$\pm$07  &  606$\pm$10 \\  
400334 & M-202082 & 00 40 55.0 & $-$13 46 31  &   3B &     24.2  &     26.2  &  112.1 &  1.944 & \phs 0.0078 & 1   &  8459 &  240 &    8463$\pm$04  &  239$\pm$05 \\  
400343 & N 217    & 00 41 33.7 & $-$10 01 18  &   0  &     48.7  &     65.9  &  270.1 &  1.875 & \phs 0.0000 & 1   &  3977 &  554 &    3976$\pm$10  &  533$\pm$14 \\  
   444 & 519-019  & 00 42 04.8 &   +36 48 16  &   3  &     22.8  &     27.1  &  218.1 &  1.337 &   $-$0.0013 & 1   & 10650 &  430 &   10644$\pm$06  &  412$\pm$08 \\  
100917 & ...      & 00 42 11.6 &   +37 00 56  &   3  &     19.2  &     16.8  &  112.7 &  6.998 & \phs 0.0350 & 1   & 10598 &  179 &   10612$\pm$08  &  223$\pm$10 \\  
400373 & M-203006 & 00 45 08.8 & $-$09 37 53  &   4  &     21.8  &     22.2  &   93.0 &  6.944 & \phs 0.0347 & 1   &  6110 &  172 &    6112$\pm$06  &  193$\pm$08 \\  
400846 & M-203014 & 00 47 25.2 & $-$09 51 10  &   4  &     16.8  &     14.6  &  115.1 &  7.491 & \phs 0.0300 & 1   &  5761 &  182 &    5761$\pm$06  &  215$\pm$08 \\  
400390 & M-203015 & 00 47 46.2 & $-$09 50 05  &   3B &     34.8  &     38.4  &  197.3 &  3.617 & \phs 0.0036 & 1   &  5767 &  418 &    5770$\pm$10  &  400$\pm$14 \\  
400387 & UA 14    & 00 47 46.5 & $-$09 53 54  &   3  &     75.1  &     63.1  &   45.8 & 13.587 & \phs 0.0242 & 1   &  1362 & \phn 98 & 1358$\pm$10  &  103$\pm$14 \\  
400848 & ...      & 00 47 51.9 & $-$09 38 46  &   3  &     14.9  &     11.4  &   74.5 &  2.565 & \phs 0.0103 & 1   &  6065 &  149 &    6065$\pm$07  &  154$\pm$10 \\  
400411 & N 268    & 00 50 09.5 & $-$05 11 37  &   5B &     31.1  &     37.9  &  118.1 &  6.949 & \phs 0.0083 & 1   &  5492 &  233 &    5494$\pm$04  &  238$\pm$06 \\  
   534 & N 280    & 00 52 30.2 &   +24 21 04  &   3  &     54.9  &     43.0  &  267.6 &  2.383 &   $-$0.0024 & 1   & 10185 &  526 &   10195$\pm$22  &  489$\pm$30 \\  
400520 & N 325    & 00 57 47.5 & $-$05 06 45  &   6  &     28.8  &     28.1  &  184.7 &  5.402 &   $-$0.0044 & 1   &  5488 &  344 &    5487$\pm$04  &  353$\pm$06 \\  
400522 & N 329    & 00 58 01.3 & $-$05 04 16  &   3  &     38.1  &     33.1  &  198.4 &  3.418 & \phs 0.0103 & 1   &  5271 &  436 &    5267$\pm$10  &  435$\pm$13 \\  
400662 & ...      & 00 59 10.3 & $-$01 53 57  &   3  &     12.8  &     13.2  &  232.2 &  4.202 &   $-$0.0271 & 1   & 15541 &  364 &   15546$\pm$07  &  386$\pm$10 \\  
   641 & N 353    & 01 02 24.5 & $-$01 57 29  &   2B &     21.8  &     21.0  &  174.5 &  4.963 &   $-$0.0099 & 1   &  4107 &  310 &    4108$\pm$08  &  325$\pm$11 \\  
410588 & M-103075 & 01 02 48.0 & $-$06 24 41  &   3  &     24.2  &     21.5  &  246.0 & 13.316 &   $-$0.0799 & 1   &  5775 &  312 &    5773$\pm$10  &  346$\pm$14 \\  
410497 & 384-060  & 01 02 48.2 & $-$01 28 24  &   3  &     18.9  &     26.2  &   96.3 &  4.158 & \phs 0.0565 & 1   &  5876 &  233 &    5876$\pm$05  &  235$\pm$07 \\  
410752 & ...      & 01 02 55.4 & $-$06 24 42  &   3  &     13.4  &      9.3  &  200.3 &  1.572 & \phs 0.0010 & 1   & 13604 &  380 &   13603$\pm$08  &  386$\pm$10 \\  
410968 & ...      & 01 03 03.7 & $-$01 42 49  &   0  &  ...\phn  &  ...\phn  &  ...\phn  &  ...\phn  &  \phn ...  & ... & 13991 &  357 &  ...\phn\phn &  \phn ... \\  
410508 & ...      & 01 03 41.2 & $-$02 20 15  &   3  &     16.0  &      9.1  &  309.8 &  1.922 &   $-$0.0006 & 1   & 21266 &  568 &   21266$\pm$31  &  576$\pm$41 \\  
410512 & M-103083 & 01 03 56.2 & $-$02 15 44  &   3  &     17.9  &     19.2  &  233.5 &  2.408 & \phs 0.0120 & 1   & 15751 &  467 &   15759$\pm$04  &  483$\pm$05 \\  
410518 & 384-064  & 01 04 15.0 & $-$01 06 51  &   3B &     15.3  &     12.2  &  111.8 &  5.999 & \phs 0.0420 & 1   &  5420 &  206 &    5409$\pm$15  &  224$\pm$21 \\  
410519 & ...      & 01 04 21.7 & $-$01 31 56  &   4  &     16.9  &     14.5  &   93.1 &  7.143 & \phs 0.0214 & 1   &  5089 &  137 &    5084$\pm$05  &  174$\pm$07 \\  
\enddata 
\end{deluxetable}
\clearpage
\end{landscape}

\end{document}